\documentclass{ws-ijgmmp}

\usepackage[colorlinks,citecolor=blue,urlcolor=blue,linkcolor=blue]{hyperref}
\usepackage{tikz}           
\usetikzlibrary{positioning}
\usepackage{xcolor}

\graphicspath{{./figs/}}

\newcommand{\mscript}[1]{{\mbox{\scriptsize \rm #1}}}
\newcommand{\gbar}{{\overline g}}
\newcommand{\rn}{{r_\mscript{n}}}

\newcommand{\VS}{{V^2_\mscript{*}}}
\newcommand{\Vgas}{{V^2_\mscript{gas}}}
\newcommand{\Vbar}{V^2_{\mscript{bar}}}
\newcommand{\Vmod}{V^2_{\mscript{mod}}}
\newcommand{\Vobs}{V^2_{\mscript{obs}}}

\newcommand{\bs}{\boldsymbol}

\newcommand{\rs}{{\rho^*}}
\newcommand{\aii}{{\tilde{a}_2}}
\newcommand{\aiii}{{\tilde{a}_3 }}
\newcommand{\eff}{{\mbox{\scriptsize eff}}}

\renewcommand{\a}{\approx}

\newcommand{\mnras}{Mont.Notices Royal Astron.Soc.}
\newcommand{\aj}{Astron. J.}
\newcommand{\apj}{Astrophysical J.}

\begin{document}

\markboth{Hernandes-Arboleda, Rodrigues, Wojnar, Toniato}{Palatini $f(R)$ gravity tests in weak field limit}

%
\catchline{}{}{}{}{}
%

\title{Palatini $f(R)$ gravity tests in the weak field limit: Solar System, seismology and galaxies}

\author{Alejandro Hernandez-Arboleda}
\address{Cosmo-Ufes \& PPGCosmo, \\ Universidade Federal do Espírito Santo, Vitória, ES, Brazil. \\
\email{alejandro.arboleda@aluno.ufes.br}}

\author{Davi C. Rodrigues}
\address{Departamento de Física \& Cosmo-Ufes \& PPGCosmo,\\  Universidade Federal do Espírito Santo, Vitória, ES, Brazil. \\ Institut für Theoretische Physik, \\Universität Heidelberg, Heidelberg, Germany.\\
\email{davi.rodrigues@ufes.br}} 

\author{Júnior D. Toniato}
\address{Departamento de Química e Física \& Cosmo-Ufes, \\ Universidade Federal do Espírito Santo, Alegre, ES, Brazil.\\
\email{junior.toniato@ufes.br}}

\author{Aneta Wojnar}
\address{Departamento de F\'isica Te\'orica \& IPARCOS, \\
Universidad Complutense de Madrid, E-28040, 
Madrid, Spain.\\
\email{awojnar@ucm.es}}

\maketitle

\begin{history}
\received{(Day Month Year)}
\revised{(Day Month Year)}
\end{history}

\begin{abstract}
Palatini $f(R)$ gravity is probably the simplest extension of general relativity (GR) and the simplest realization of a metric-affine theory. It has the same number of degrees of freedom as GR and, in vacuum, it is straightforwardly mapped into GR with a cosmological constant. The mapping between GR and Palatini $f(R)$ inside matter is possible but at the expense of reinterpreting the meaning of the matter fields. The physical meaning and consequences of such mapping will depend on the physical context. Here we consider three such cases within the weak field limit: Solar System dynamics, planetary internal dynamics (seismology), and galaxies. After revising our previous results on the Solar System and Earth's seismology, we consider here the possibility of $f(R)$ Palatini as a dark matter candidate. For any $f(R)$ that admits a polynomial approximation in the weak field limit, we show here, using SPARC data and a recent method that we proposed, that the theory cannot be used to replace dark matter in galaxies. We also show that the same result applies to the Eddington-inspired Born-Infeld gravity. Differently from the metric $f(R)$ case, the rotation curve data are sufficient for this conclusion. This result does not exclude a combination of modified gravity and dark matter.
\end{abstract}


\section{Introduction}	

The $f(R)$ gravity models are simple and natural extensions of GR. Three large classes of $f(R)$ gravity are the metric, the Palatini, and the hybrid cases \cite{Capozziello:2010zz, Capozziello:2011et, Olmo:2011uz, Capozziello:2015lza,tamanini2013generalized}. The Palatini case starts with a bold generalization: the affine connection is independent of the metric. This suggests an increase in the number of degrees of freedom concerning GR. However, the field equations demand a relation between the metric and the connection, leading to two degrees of freedom and becoming identical to GR with cosmological constant in vacuum \cite{Ferraris:1992dx, Rodrigues:2018ioe}. Inside matter, the correspondence to GR will in general depend on the definition of the physical system (e.g., \cite{TeppaPannia:2016vsb, Wojnar:2017tmy, Toniato:2019rrd,jimenez2018born,Wojnar:2022dvo,Benito:2021ywe,Kalita:2022trq}).

Due to the equivalence between $f(R)$ Palatini and GR in vacuum, it is evident that the parameterised post-Newtonian (PPN) formalism in the Eddington-Robertson-Schiff approach \cite{1972gcpa.book.....W} will pose no constraints; this since this approach is based on a vacuum solution. On the other hand, the PPN test due to Will-Nordtvedt (WN) \cite{Will:2014kxa} is not based on Schwarschild solution or an extension of it, moreover it tests 10 PPN parameters in its standard form, instead of two from the Eddington-Robertson-Schiff approach.  Nonetheless, in \cite{Toniato:2019rrd} we have shown in detail that the WN PPN formalism cannot put constraints on $f(R)$ Palatini. The latter result depends on the redefinition of the pressure and internal energy inside matter, but such redefinitions cannot be tested in the context of the Solar System PPN application. 

The PPN result above should not be seen as a statement that $f(R)$ Palatini will always yield the same result of GR, independently on the physical context. Indeed, in \cite{Kozak:2023axy} some of us used seismology data from Earth and find a constraint on the theory, which is here revised.

It is well known that some specific metric $f(R)$ theories can yield good results for galaxy rotation curves without dark matter (e.g., \cite{Capozziello:2006ph, Capozziello:2012ie, Salucci:2014oka,  Naik:2019moz}). We test here if it is possible for Palatini to do something similar. This is done using 122 galaxies form the SPARC dataset \cite{2016AJ....152..157L}. To this end, we use an approach that some of us developed recently \cite{Rodrigues:2022oyd}. 

In the next section, we review $f(R)$ Palatini gravity in the weak field limit. Section \ref{sec:ppn} is devoted to the PPN review, Sec.~\ref{sec:seismology} for the seismology review, and in Sec.~\ref{sec:galaxies} we present our main results. In \ref{appA} we briefly explain the connection between the Palatini weak field limit and Eddington-inspired-Born-Infeld (EiBI) theory \cite{Banados:2010ix}. This is relevant since some of our results only depend on the weak field limit of $f(R)$ Palatini, these results are therefore automatically valid for EiBI as well.

\section{Palatini $f(R)$ and its weak field limit: a brief review} 
 
Here we review key aspects of Palatini $f(R)$ that are relevant for this work. Our notation is close to the one we used in \cite{Toniato:2019rrd}. Extended reviews on Palatini $f(R)$ gravity can be found in Refs.~\cite{Sotiriou:2008rp, DeFelice:2010aj, Olmo:2011uz}.  

The (torsionless) Palatini $f(R)$ gravity action, in the presence of a matter action $S_{\mscript{m}}$ that depends on the matter fields $\Psi$, reads,
\begin{equation}
	S[g,\Gamma, \Psi] = \frac{1}{2\kappa}\int f\big(R(\Gamma,g)\big) \sqrt{-g} \, d^4x + S_{\mscript{m}}[g,\Psi] \,. \label{actionP}
\end{equation}
In the above $\kappa$ is a constant, $g_{\mu \nu}$ is the spacetime metric, ${\Gamma}^{\lambda}_{\mu\nu}$ is an independent and torsionless affine connection. That is, in the action, $\Gamma_{\mu \nu} ^\lambda$ enters as an independent field from the spacetime metric and ${\Gamma}^{\lambda}_{\mu\nu} = {\Gamma}^{\lambda}_{\nu \mu} $.
The Riemann tensor is such that, for any vector field $A^\lambda$, $R^\lambda_{ \sigma \mu \nu} A^\sigma= [ \nabla_\mu, \nabla_\nu] A^\lambda$, where $\nabla_\mu$ is the covariant derivative with respect to the affine connection ${\Gamma}^{\lambda}_{\mu\nu}$. The Ricci tensor and Ricci scalar are respectively $R_{\mu \nu}(\Gamma) \equiv R^\lambda_{\mu \lambda \nu}(\Gamma)$ and $R \equiv g^{\mu \nu} R_{\mu \nu}(\Gamma)$. Let us notice that in the case of Palatini gravity the matter fields couple only to the spacetime metric, not to the connection $\Gamma$. 

The field equations derived from $g_{\mu \nu}$ and $\Gamma^\mu_{\nu \lambda}$ variations are, respectively,
\begin{align}
	&f'(R)R_{\mu\nu}-\dfrac{1}{2}\,f(R)g_{\mu\nu}=\kappa\, T_{\mu\nu}\,,\label{feq}\\[.1in]
	&\nabla_\lambda\left(\sqrt{-g}f'(R)g^{\mu\nu}\right)=0\,.\label{cov}
\end{align}
The prime on $f(R)$ means the derivative with respect to $R$.  $T_{\mu \nu}$ is the energy-momentum tensor, which is defined by \cite{Wald:1984rg}
\begin{equation}
	T_{\mu \nu} = -\frac{2}{\sqrt{- g}}\frac{\delta S_\mscript{m}}{\delta g^{\mu \nu}} \, .
\end{equation}

Equation \eqref{feq}, together with the matter field equations, leads to $\nabla^\mu T_{\mu \nu} \not=0$ (where we recall that $\nabla_\mu$ is the covariant derivative built from the connection $\Gamma^\mu_{\nu \lambda}$). This is expected since the action variation with respect to $\Psi$, together with diffeomorphism invariance of $S_\mscript{m}$, implies that\footnote{The appendix of Ref.~\cite{Fabris:2020uey} presents further general considerations about $T_{\mu\nu}$ conservation, in a modified gravity context.}
\begin{equation}
	\nabla_{\!C}^\mu T_{\mu \nu} =0 \, , \label{consv}
\end{equation}
where $\nabla_{\!C}^\mu$ is the covariant derivative with respect to the connection
\begin{equation}
	C_{\mu \nu}^\lambda = \frac 12 g^{\lambda \sigma}\left(g_{\mu \sigma, \nu} + g_{\sigma \nu, \mu} - g_{\mu \nu, \sigma}\right) \, .
\end{equation}
In the above, a comma is used to denote a partial derivative. In particular, this implies that test particles will follow geodesics from the connection $C_{\mu \nu}^\lambda$, not $\Gamma_{\mu \nu}^\lambda$. 

From the trace of eq.~\eqref{feq}, 
\begin{equation}
	f'(R) R - 2 f(R) = \kappa T \,, \label{tracefeq}
\end{equation}
which shows that $R$ can be algebraically expressed as a function of $T$. This is in  contrast to the metric $f(R)$ case. 

In order to proceed with the weak field limit, and similarly to \cite{Toniato:2019rrd}, we consider the following  $f(R)$ case, 
\begin{equation}
	f(R) = R + \alpha R^2 + O(R^3) \,, \label{fRexp}
\end{equation}
where $\alpha$ is a constant.
With the above, we are assuming that $f(R)$ can be approximated by an analytical function about $R=0$, with $f(0)= O(R^3)$. We stress that the condition  $f(0)=O(R^3)$ is more general than $f(0) =0$. This condition is natural and, considering the PPN context, it is necessary. It is necessary to guarantee asymptotic flatness, otherwise PPN in its standard form cannot be applied. Moreover, we are only assuming that a cosmological constant-like term, if present, is not relevant for the considered weak field system. Even if the complete $f(R)$ function may be non-analytical, the approximation above is valid for any $f(R)$ without singularities about $R=0$. In particular, this approximation is not valid for models of the type $f(R) = 1/R + O(R)$, which are commonly prone to strong physical constraints (e.g., \cite{kainulainen2007interior}).

Hence, eq.~\eqref{tracefeq} implies that
\begin{equation}
	R = - \kappa T + O(R^3). \label{RTrelation}
\end{equation}

It is convenient to express the connection $\Gamma_{\mu \nu}^\lambda$ as a function of a ``metric'', and indeed this is well known to be  possible.  Equation \eqref{cov} points out that $\Gamma^\lambda_{\mu \nu}$ can be written as a function of a tensor $\gbar_{\mu \nu}$ that is conformally related to the spacetime metric $g_{\mu \nu}$; indeed, let
\begin{equation}\label{gbar}
	\gbar_{\mu\nu}=f'(R)\,g_{\mu\nu} \, ,
\end{equation}
and hence the independent connection results as 
\begin{equation}
 	\Gamma_{\mu \nu}^\lambda = \frac 12 \gbar^{\lambda \sigma}\left(\gbar_{\mu \sigma, \nu} + \gbar_{\sigma \nu, \mu} - \gbar_{\mu \nu, \sigma}\right) \, .
\end{equation}

As usual for the weak field limit, the background is taken as Minkowski $\eta_{\mu \nu}$,
\begin{equation}
	\gbar_{\mu \nu} = \eta_{\mu \nu} + \overline h_{\mu \nu}\, .
\end{equation}

Using eq.~\eqref{RTrelation}, up to the first nontrivial order, the field equation \eqref{feq} does not depend on $\alpha$ and it is equivalent to the Einstein field equation for the metric $\gbar_{\mu \nu}$, coupling constant $\kappa$ and energy-momentum tensor $T_{\mu \nu}$.

For the three systems considered in this work, the relevant energy-momentum tensor is that of a perfect fluid,
\begin{align}\label{perf-fluid}
	T^{\mu\nu}=\left(\rho+\rho\Pi+p\right)u^{\mu}u^{\nu}+pg^{\mu\nu}\,,
\end{align}
where $\rho$ is the mass density, $\Pi$ is the fluid's internal energy per unity of mass, $p$ is the pressure and $u^{\mu}=u^{0}(1,\bs v)$ is the four-velocity of the fluid.
Since the fluid particles are not assumed to move at relativistic velocities, we consider $\rho \gg p$ and thus, as a first approximation, $T\approx - \rho$. 

Therefore, using $\overline h_{ 0 0} = - 2 \overline \Phi$,
\begin{equation}
	\nabla^2 \overline \Phi \approx \frac{\kappa}{2} \rho \, . \label{PoissonBar}
\end{equation}

The difference with respect to GR comes from the Euler equation, which is derived from eq.~\eqref{consv} and depends on the connection $C_{\mu \nu}^\lambda (g)$, instead of $\Gamma^\lambda_{\mu \nu} (\gbar)$. This implies that test particles will accelerate due to a gravitational potential $\Phi$, with
\begin{equation}\label{expand-g}
g_{\mu \nu} = \eta_{\mu \nu} + h_{\mu \nu},
\end{equation}
and $h_{ 0 0} = - 2  \Phi$. The Poisson equation becomes 
\begin{equation} \label{polatiniPotential}
	\nabla^2 \Phi \approx \frac{\kappa}2 \left( \rho + 2 \alpha \nabla^2\rho \right) \, .
\end{equation}

The above is the modified Poisson equation implied by Palatini $f(R)$ gravity in the weak field regime. This equation was derived in \cite{Toniato:2019rrd}. We also point out that Eddington-inspired Born-Infeld (EiBI) gravity \cite{Banados:2010ix} has precisely the same weak field limit. This is shown in \ref{appA}. We focus this work on $f(R)$ Palatini, but clearly several of our results can be immediately extended to EiBI gravity.

\section{Post-Newtonian analysis for Solar System tests} \label{sec:ppn}

The study of the post-Newtonian (PN) equations of motion of a gravitational theory is mandatory to confront Solar System observational data and theoretical predictions. For torsionless Palatini $f(R)$ models, whose Lagrangian is analytic about $R=0$, PN analysis has been made in Ref. \cite{Toniato:2019rrd} under the optic of the parametrized post-Newtonian (PPN) formalism \cite{Will:1993ns}. In this section, we review their main results.

It is not possible to apply the PPN formalism to a fully arbitrary $f(R)$ function. But, for the large and relevant class of functions that can be \textit{approximated} by a polynomial expansion in the weak field limit, it can be applied. For the PN analysis, it is relevant to require the knowledge of the $f(R)$ approximation up to the $R^3$ term, hence let
\begin{equation}\label{poly}
f(R)=R + a_2 R^2 + a_3 R^3 + O(R^4) ,
\end{equation}
with the coefficients $a_2$ and $a_3$ being constants. This notation is the same used in \cite{Toniato:2019rrd}. We stress that the $\alpha$ introduced in eq.~\eqref{fRexp} is identical to $a_2$ ($\alpha = a_2$). In this PPN context, we  use $a_2$, once $\alpha_i$ symbols usually represent PPN parameters.

We work with the same assumptions as in the standard PPN formalism, therefore, asymptotic flatness demands no constant term in the expansion of $f(R)$, while the coefficient for the linear term can be made equal to 1 due to the arbitrariness of the theory's coupling constant. The Newtonian limit must also be assumed, and this is possible with the expansion \eqref{poly}.

To have complete PN equations of motion, it is only necessary to know  $g_{00}$ up to fourth-order terms, $g_{0i}$ up to third order and $g_{ij}$ to second order. After solving field equations, the expanded metric components will be given by
\begin{align}
g_{00} \a & \ -1 + 2 U + 2\Psi - 4\aii \Phi_{P} + 3 \aiii  {\rho^*}^2   \notag \\
	& \ + \aii \rs\!\left( 2 - 10U +  v^2\!  - 2 \Pi  + 6p/\!\rs \! -  2 \aii \rho^* \right)\!,\label{g004} \\[1ex]
	g_{0i} \a& \ -4V^{i},\label{g0i}\\[1ex]
	g_{ij} \a & \ \delta_{ij}+2\left(U -\aii\rs \right)\delta_{ij}.\label{gij}
\end{align}
where we have worked with units where $G=c=1$, taking $\kappa=8\pi$ and we defined $\aii=8\pi a_2$ and $\aiii=64\pi^{2}a_{3}$. The $\rs\equiv{u^{0}}\sqrt{-g}\,\rho$ is the conserved fluid's rest mass density. Also, $U$ is the negative of the Newtonian potential, $\Psi$ groups all the fourth-order standard PPN potentials and $\Phi_{P}$ is a particular potential for Palatini $f(R)$ gravity. All these potentials are defined in terms of the conserved density. 

The first issue related to the expression \eqref{g004} is that in the absence of matter, it does not recover the equivalent expansion of GR. This is so because the potential $\Phi_P$ does not vanish in vacuum regions (the remaining terms in $g_{00}$ with $\rs=p=0$ coincide with the ones obtained in GR). Hence, in order to obtain the equivalence between Palatini $f(R)$ and general relativity in vacuum, one must construct a redefinition of matter quantities, namely
\begin{equation}
	\rho^* \Pi_\eff + 3 p_\eff = \rho^* \Pi + 3 p - 2 \aii {\rho^*}^2 \, . \label{PipEff}
\end{equation} 
We will further comment on the implications of such redefinition.

It would be tempting to obtain all the PPN parameters from metric components \eqref{g004}--\eqref{gij}. However, there are several terms in these expressions that are outside the standard formalism parametrization. It is common to see in the literature analysis exploring definitions of effective PPN parameters that would encompass the PN metric modifications brought by an alternative theory and use them to constraint-free parameters. Notwithstanding, we disagree with this method of work since all the PPN parameters have their physical meaning (i.e. the precise relation to the observational phenomena) constructed with the assumption that they are constant in time and space. The correct approach should be to reanalyse conservation laws and equations of motion in order to verify if and how the new PN terms affect the original PPN parameters.

In the specific case of Palatini $f(R)$,  the trajectory of electromagnetic waves (or light in particular) can be understood directly.  Since all the experiments used to constrain the light motion do not depend on a given medium, and only include second-order metric perturbations, such electromagnetic experiments cannot infer any difference with respect to GR. Therefore, it is possible to say that $\gamma=1$ within Palatini $f(R)$ gravity. This is in contrast with Ref.~\cite{Olmo:2005zr}: the reason for the difference comes from non-asymptotically flat effects which should not be considered in PPN, unless a proper explanation of the validity of the method is included.

From equation \eqref{consv}, and considering the PN metric components \eqref{g004}-\eqref{gij}, one can extract three conservation statements. For $\mu=0$ the leading terms recover the continuity equation for $\rs$ and a conserved mass $m$ is found. The total energy $E$ which is conserved in post-Newtonian Palatini expansion is obtained from the next leading order and it is the same as the one derived in GR with a source whose effective internal energy is given by $\Pi_\eff=\Pi - \aii \rs/2$. Moreover, with the definition of the gravitational mass as $M=m+E$, both theories also agree on this quantity, which is consistent with their equivalence in vacuum.

The previous redefinition of the fluid's internal energy, together with condition \eqref{PipEff}, fixes an expression for $p_\eff$ also that is justly the pressure term appearing at the Newtonian-order of Euler's equation. The latter is derived from the leading order terms of the case $\mu = i$ in eq.~\eqref{consv}, with the next order determining the post-Newtonian conserved momentum in Palatini $f(R)$. Thus, one can note that $p_\eff$ is the relevant pressure in the sense that it is the one giving a force per unity of area that is locally generated by the fluid. This is the usual method to macroscopically measure pressure. Within this approach, the effective pressure and internal energy are the important information from the fluid.
One could also interpret that a fluid's pressure or its internal energy may be determined from microphysics principles, which would put into test the aforementioned redefinitions (see Ref.~\cite{Toniato:2019rrd} for a more detailed discussion).

Notwithstanding, Palatini $f(R)$ does not violate total conservation of energy and momentum in the PN regime and, within the context of the PPN formalism, this indicates that the parameters $\zeta_1,\zeta_2,\zeta_3,\zeta_4$ and $\alpha_3$ did not have their physical meaning altered by the presence of the $\Phi_P$ potential in the PN expansion of the metric. In conclusion, all these five parameters are zero in Palatini $f(R)$ gravity, which is consistent with the observational bounds.

The remaining PPN parameters to be determined are $\beta$, $\alpha_1$ and $\alpha_2$. Using the previous results, this task is completed through the analysis of the equation of motion of massive bodies. The procedure is to split the fluid description into $N$ well-separated bodies and describe their motion in terms of centre-of-mass positions. The details are described in Ref. \cite{Toniato:2019rrd}.

The final results for a body $A$ acceleration can be written as
\begin{eqnarray}
\bs a_{\scriptscriptstyle A} &=& \sum_{\scriptscriptstyle B\neq A}\dfrac{M_{\scriptscriptstyle B}}{r_{\scriptscriptstyle AB}^{2}}\bigg\{\!
	 \bigg[1+ v_{\scriptscriptstyle A}^{2}-4(\bs v_{\scriptscriptstyle A}\!\cdot\!\bs v_{\scriptscriptstyle B})+2v_{\scriptscriptstyle B}^{2}\,- \dfrac{3}{2}(\bs n_{\scriptscriptstyle AB}\!\cdot\!\bs v_{\scriptscriptstyle B})^{2} -\dfrac{5M_{\scriptscriptstyle A}}{r_{\scriptscriptstyle AB}} -\dfrac{4M_{\scriptscriptstyle B}}{r_{\scriptscriptstyle AB}} \bigg] \bs n_{\scriptscriptstyle AB} \notag\\[1ex]
	 && \ \ - \ \Big[ \bs n_{\scriptscriptstyle AB}\cdot (4\bs v_{\scriptscriptstyle A}-3\bs v_{\scriptscriptstyle B}) \Big] (\bs v_{\scriptscriptstyle A} -\bs v_{\scriptscriptstyle B})\,+ \dfrac{7}{2}\sum_{C\neq A,B}\!\! M_{\scriptscriptstyle C}\dfrac{r_{\scriptscriptstyle AB}}{r_{\scriptscriptstyle BC}^{2}}\,\bs n_{\scriptscriptstyle BC} \ +  \label{apn} \\[1ex]
	 && \ \ - \sum_{\scriptscriptstyle C\neq A,B}M_{\scriptscriptstyle C}\left[ \dfrac{4}{r_{\scriptscriptstyle AC}} +\dfrac{1}{r_{\scriptscriptstyle BC}} - \ \dfrac{r_{\scriptscriptstyle AB}}{2r_{\scriptscriptstyle BC}^{2}}(\bs n_{\scriptscriptstyle AB}\cdot\bs n_{\scriptscriptstyle BC}) \right]\bs n_{\scriptscriptstyle AB} \nonumber
	  \bigg\} \, .
\end{eqnarray}
In the above expressions, we use the definitions $M_{\scriptscriptstyle A}=m_{\scriptscriptstyle A}+E_{\scriptscriptstyle A}$, $\bs r_{\scriptscriptstyle AB}=\bs r_{\scriptscriptstyle A} -\bs r_{\scriptscriptstyle B}$, $r_{\scriptscriptstyle AB}=|\bs r_{\scriptscriptstyle AB}|$ and $\bs n_{\scriptscriptstyle AB}=\bs r_{\scriptscriptstyle AB}/r_{\scriptscriptstyle AB}$.

These expressions for the accelerations are \textit{identical} to the corresponding GR expressions \cite{poisson_will_2014} (this is evident also by the absence on either $a_2$ or $a_3$). Thus, one can compare the term $\bs a_{\scriptscriptstyle A}^{\mbox{\tiny PN}}$ with the general PPN formula for the post-Newtonian acceleration and conclude that $\beta=1$ and $\alpha_1=\alpha_2=\xi=0$. Therefore, the values of all the PPN parameters in Paltini $f(R)$ are the same as GR. Moreover, the physical meaning of all parameters are not changed by the presence of a non-standard PPN potential in the metric expansion. In conclusion, all the bounds on the PPN parameters not only can be safely used for Palatini $f(R)$ but they also are all in complete agreement with observations. It should be noted that this result does not imply complete equivalence between Palatini $f(R)$ and GR, just that they cannot be differentiated within the PN test frameworks.

Although the last results  guarantee that Palatini $f(R)$ is in agreement with all Solar System tests, it also points out that the theory can only be constrained in systems where further knowledge of the matter source is possible. As we will see next, planetary seismology should allow us to move forward to this task.


\section{Planetary Physics} \label{sec:seismology}
Recent advancements in seismographic tools and laboratory experiments have significantly increased our knowledge about the interior of our planet \cite{poirier2000introduction,dziewonski1981preliminary,kustowski2008anisotropic}. This includes the study of iron's properties and behaviour under extreme temperatures and pressures in the Earth's core \cite{merkel2021femtosecond}. Furthermore, new neutrino telescopes are providing valuable information on the density, composition, and abundance of light elements in the outer core, reducing uncertainties related to the Earth's core characteristics \cite{donini2019neutrino,winter2007neutrino,van2021probing}.

There is a small but still notable impact of Palatini gravity effects on planetary physics. The correction term in the Poisson equation \eqref{polatiniPotential} (and in the resulting hydrostatic equilibrium equation, see below) has not only an effect on the early \cite{Wojnar:2022ttc} and late \cite{Wojnar:2021xbr} evolution of the giant (exo-)planets, but also on the interiors of the solid, Earth-like ones \cite{Kozak:2021zva,Kozak:2021fjy}. 

Due to that fact, since modified gravity affects the densities and thicknesses of the layers \cite{Kozak:2021ghd} in solid planets, we have a chance to constrain gravitational models, especially in the light of more accurate data about the Solar System planets and the Earth in particular. However, in order to begin constraining theories, we need an Earth model as a reference. The most commonly used global seismological Earth model is the Preliminary Reference Earth Model (PREM), which provides pressure, density, and elastic moduli profiles as functions of depth \cite{dziewonski1981preliminary}, see the figure 1 in \cite{Kozak:2023axy}. This model is based on velocity-depth profiles given by the travel-time distance curves for seismic waves and periods of free oscillations. The missing element to determine these profiles is a hydrostatic equilibrium equation which in case of PREM is the Newtonian one
   \begin{equation}
  \frac{dP}{dr}= -\rho \frac{GM(r)}{r^2}),
    \end{equation}
thus it does depend on a model of gravity. Since we do not have in our disposal a gravity-independent model of Earth yet, one assumes that PREM is an accurate model of our planet. Being equipped with the mass and moment of inertia of our planet, we will briefly discuss how one can use the Earth's seismic data to constrain a given gravitational proposal \cite{Kozak:2023axy}. To do so, we further assume that the planet is spherical-symmetric and that we are dealing with adiabatic compression - that is, there is no heat exchange between the layers such that we can neglect the temperature variation term in the hydrostatic equilibrium equation:
\begin{align}
    \frac{d\phi}{dr}&=-\rho^{-1}\frac{dP}{dr},.
\end{align}
It can be derived from \eqref{polatiniPotential} for Palatini $f(R)$:
    \begin{equation}\label{hydro}
  \frac{dP}{dr}= -\rho\left( \frac{GM(r)}{r^2}+\alpha\kappa\rho'(r)\right) =: -\rho g_\text{eff},
    \end{equation}
 where the pressure $P$ and density $\rho$ are functions of the radial coordinate $r$ while mass $M(r)$ is given by
\begin{equation}\label{mass}
    M(r)= 4\pi \int_0^r r^2\rho(r) dr.
\end{equation}
Moreover, another assumption is made on the planet's layers: they are radially symmetric shells with the density jump between the inner and outer core $\Delta\rho=600$, central density $\rho_c=13050$ and density at the mantle's base $\rho_m=5563$ (in kg/m$^3$). The outer layers (starting from the upper mantle) are described by the Birch law $\rho = a + b v_p$, in which $a$ and $b$ are parameters depending on the mean atomic mass of the material in the upper mantle \cite{dziewonski1981preliminary}. The velocity $v_p$ is the longitudinal elastic wave and it appears in the seismic parameter $\Phi_s$ \cite{poirier2000introduction}
    \begin{equation}
        \Phi_s= v_p^2 - \frac{4}{3} v_s^2,
    \end{equation}
    together with the transverse elastic one $v_s$.
Since the seismic parameter is related to the bulk modulus (incompressibility of a given material)
\begin{equation}
    K= \frac{dP}{d \mathrm{ln}\rho}
\end{equation}
as $ \Phi_s=  K/{\rho}$, it carries the information on the properties of matter, that is, the equation of state,
\begin{equation}
     \Phi_s=  \frac{dP}{d\rho}.
\end{equation}
Therefore, we can rewrite \eqref{hydro} as
  \begin{equation}\label{hydro2}
        \frac{d\rho}{dr} =  -\rho g_\text{eff}  \Phi_s^{-1}.
    \end{equation}
Mass \eqref{mass} and moment of inertia ($R$ is Earth's radius)
\begin{equation}
    I= \frac{8}{3}\pi \int_0^R r^4\rho(r) dr,
\end{equation}
are our constraints, given by observations with a high accuracy \cite{luzum2011iau,chen2015consistent}. 

The solution of the hydrostatic equilibrium equation \eqref{hydro2} with the mass equation \eqref{mass} and seismic data (the longitudinal and transverse elastic waves $v_p$ and $v_s$ given by \cite{dziewonski1981preliminary}) provide the density profile. Because of the modifications in \eqref{hydro2}, one expects to obtain different density profiles \cite{Kozak:2021ghd,Kozak:2023axy} than the ones given by the PREM model. Indeed, choosing a few values of the parameter $\alpha$, one obtains slightly different profiles with respect to the Newtonian one, see figures in \cite{Kozak:2021zva}. 

The effects of gravity models on a planet's mass and moment of inertia are given by calculating errors for fixed parameter values $(\rho_c, \rho_m, \Delta \rho)$, see figure 4 in \cite{Kozak:2023axy}. The results show that the deviations exceed the 2$\sigma$ level when the parameter $\alpha$ is around $10^9$ m$^2$, similar to the upper limits in other works \cite{olmo2005gravity,banerjee2017constraints}. The planet's mass is slightly more sensitive to changes in $\beta$ than its moment of inertia due to the outermost layers' significant contribution to the moment of inertia calculation (which is assumed to satisfy Birch law). However, even small changes in the planet's core structure and composition can significantly affect its moment of inertia, as we could see. Using this property and  simple analysis summarised here, one can constrain Palatini $f(R)$ ($\alpha\lesssim 10^9 \text{m}^2$) and Eddington-inspired Born-Infeld gravity\footnote{See \ref{appA}.} with the seismic data.

\section{Galaxies} \label{sec:galaxies}

\subsection{General considerations}
Here we consider the possible effects of $f(R)$ Palatini on galaxy dynamics. In particular, we test the possibility of replacing dark matter with gravitational effects. To this end, instead of doing a hard-to-implement and, in this context, not particularly informative analysis of galaxy-by-galaxy fits, we use the recently proposed Normalized Additional Velocity (NAV) method \cite{Rodrigues:2022oyd}.

For an axisymmetric system in the weak field limit \eqref{polatiniPotential} with cylindrical radial coordinate $r$ and density $\rho(r,z)$,   the circular velocity in the plane $z=0$ is
\begin{align}
	\frac{V^2}{r} &= \partial_r \Phi  \nonumber \\
	& = \partial_r \overline \Phi + \kappa \alpha  \partial_r \, \rho \, . \label{V2Palatini}
\end{align}

As revised in Sec.~\ref{sec:ppn}, some of us \cite{Toniato:2019rrd}, in a post-Newtonian context, found that there are no differences between Palatini $f(R)$ and GR considering the Solar System dynamics (apart from a map between fluid quantities that cannot be tested in that context). There is a Newtonian order correction to gravity, given by $\kappa \alpha  \partial_r \, \rho$, but this correction is irrelevant for the center of mass trajectory of a body. Thus Newtonian gravity fixes $\kappa = 8 \pi G$. However, this $\rho$ dependent correction is, in general, relevant for the internal dynamics of fluids, in particular for the stability of planets or galaxies \cite{Toniato:2019rrd}. We consider here this development to be galaxies. 

Before proceeding, there is still a relevant consideration that needs to be fixed: the fluid description. We assume here that a galaxy is a fluid that can be decomposed into stellar and gaseous parts. We will not consider here the dark matter component, since we look at the possibility of replacing it with Palatini $f(R)$ effects. These two considered parts are assumed to have negligible pressure, in the sense $p\approx 0$. Therefore, eq.~\eqref{PoissonBar} is here interpreted as the Newtonian Poisson equation that is usually applied to galaxies, with $\kappa = 8 \pi G$, and $\alpha$ parameterises Newtonian gravity deviations.

\subsection{Thin disks and $f(R)$ Palatini}

Since disk galaxies have thin disks, it is important to verify if this structure of thin disks is compatible with Palatini gravity without dark matter.  In particular, if the approximation of razor-thin disks \cite{0691084459} (i.e., disks with negligible thickness, $\rho(r,z) = \Sigma(r) \delta(z)$) is sensible within Palatini. This analysis is sufficient for finding that $\alpha < 0$, as it will be shown.

For an axisymmetric disk,  with  $z=0$ as the plane of the disk, the Newtonian acceleration $\mathbf{\bar  a} = -\boldsymbol{\nabla} \Phi$ has a non-null component that is perpendicular to the disk at $z\not=0$. 

From eq.~\eqref{polatiniPotential}, one sees that the $z$-component acceleration is given by
\begin{equation}
  a_z = \bar a_z + 8 \pi G \alpha \partial_z \rho \, ,
\end{equation} 
where $a_z = (\mathbf{\bar a})_z$ is the Newtonian $z$-component acceleration.  Hence, outside the disk the Palatini contribution is null, that is $\mathbf{a} = \mathbf{\bar a}$. In the disk, the acceleration becomes infinite if $\rho \propto \delta(z)$. 
If this force is towards the disk, no matter could escape the disk and the disk is stable, this corresponds to $\alpha < 0$. Clearly, the disk is unstable for $\alpha > 0$. Hence, there is no qualitative issue with using razor-thin disks in Palatini gravity with $\alpha < 0$. In general, disks with finite thickness would be more stable in Palatini than in Newtonian gravity.

Another issue is the disk thickness influence on the circular velocity within Palatini $f(R)$ gravity. For Newtonian gravity, there is an influence, but for the Palatini non-Newtonian contribution there is nothing more. Indeed, for $\rho(r,z) = \Sigma(r) Z(z)$, which is how disk thickness is commonly modelled, the non-Newtonian Palatini contribution to the rotation curve is the same, independently of the details of the function $Z(z)$. Since it only depends locally on the density, at $z=0$ it only depends on the constant $Z(0)$.  

In order to understand if Palatini $f(R)$ gravity can somehow replace dark matter in galaxies, the standard galaxy-by-galaxy fit approach requires the knowledge of the mass distribution of several galaxies. If it is known, then it will be necessary to solve the modified Poisson equation \eqref{polatiniPotential} for each one of them, while taking into account the observational errors on the circular velocity, mass-to-light ratios and ideally on the inclination and distance as well. After that, all the results should be put together and analysed if the systematics of the individual galaxies are reasonable in some sense. A very simple and common way to proceed is to compute the reduced chi-squared of each one of the galaxies. Nonetheless, this demands a significative amount of work that, by itself, does not clearly elucidate if the rotation curve contribution from $f(R)$ Palatini can, on average, mimic the expected non-Newtonian contribution. It is also important to stress that many well-known galaxy samples do not provide a detailed and complete profile of the baryonic mass distribution.  On the other hand, the circular velocity contributions from the gas and stellar parts are always provided (which is sufficient information for testing dark matter profiles, but not modified gravity).

One can criticize the replacement of dark matter by Palatini gravity by other arguments outside the scope of galaxy rotation curves, for instance by considering gravitational lensing or the CMB. Nonetheless, our purpose here is to study the effects of galaxy rotation curves alone, both because understanding the issues in this context may lead to new ideas, and to illustrate the usage of this recent method within a different model.

\subsection{Normalized additional velocity definition}
In \cite{Rodrigues:2022oyd} we introduced a method (the normalized additional velocity, NAV) to deal with dark matter or modified gravity effects on galaxy rotation curves. This method focuses on evaluating the shape of the dark matter or non-Newtonian contribution, instead of the amplitude. What is tested is already relevant, and it is sufficient for showing that the non-Newtonian contribution from Palatini gravity is far from sufficient to replace dark matter. Also,  the results are independent of whether $\alpha$ is constant or changes from galaxy to galaxy. The issue of Palatini effects together with dark matter models is not here considered. One would first need to address what would be the dark matter halo profile in such a case.

The observational additional velocity is defined as \cite{Rodrigues:2022oyd}\footnote{See also \cite{McGaugh:2006vv} for a different approach that uses essentially the same definition.}
\begin{equation}
	\Delta \Vobs \equiv \Vobs - \Vbar  \, . \label{DVobs}
\end{equation}
and the model additional velocity by
\begin{equation}
	\Delta \Vmod \equiv \Vmod - \Vbar \, .
\end{equation}
In the above, $V_\mscript{obs}$ is the observational rotation curve (RC), $V_\mscript{mod}$ is the model RC and $V_\mscript{bar}$ is the expected baryonic RC (which is essentially inferred from the stars and the hydrogen surface brightness). To be more explicit, 
\begin{equation}
	\Vbar =  \VS  + \Vgas \, ,
\end{equation}
where $\VS$ and $\Vgas$ are the squared circular velocities of the stellar disk  and the gas (atomic hydrogen and helium). The mass-to-light ratios are included in these quantities. In general, it is common to divide the stellar contribution into disk and bulge. Here we will only consider galaxies that do not have a relevant bulge.

These squared velocities need not be positive since they are not truly the square of a given physical velocity, they are actually the radial acceleration component multiplied by $r$ (this is a common convention).

\begin{figure}
    \centering
	\begin{tikzpicture}
  		\node (img1)  {\includegraphics[width=0.70\textwidth] {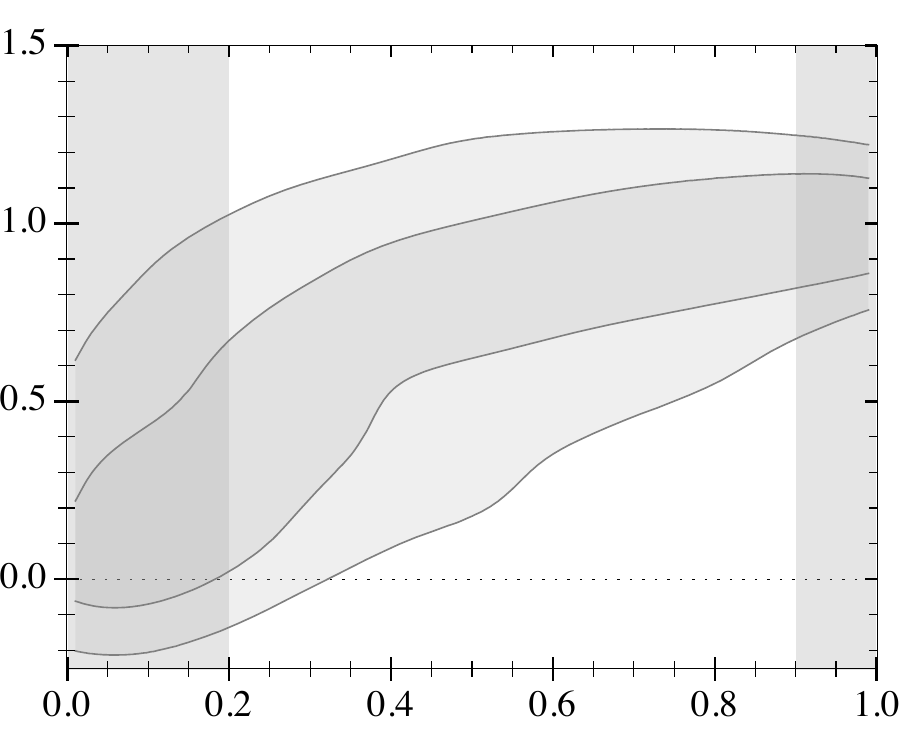}};
		\node[below=of img1, node distance=0cm, yshift=1.1cm, xshift=0.3cm, font=\color{black}] {\large $\rn$};
		\node[left=of img1, node distance=0cm, rotate=90, yshift=-0.7cm, xshift=0.6cm] {\large $\delta V^2$};
	\end{tikzpicture}
	\caption{The observational NAV plane from SPARC data. It is inferred from eq.~\eqref{NAVdef} and 122 SPARC galaxies (it uses the same quality cuts used to derived the radial acceleration relation, reducing the sample from 175 to 153,  and neglects galaxies with a relevant bulge). The two contours show the densest data regions: the densest region with 68\% of the data in dark grey (1$\sigma$ region), and the densest region which includes about 95\% of the data is in light grey (2$\sigma$ region). These regions are not confidence regions, they display the distribution of observational data. The two vertically greyed regions display the regions with larger uncertainty which are not considered in any NAV analysis.}
	\label{fig:plotNAVbulgelessObs} 
\end{figure}

The method is based on the normalized additional velocity (NAV), denoted by $\delta V^2$,  as a function of the normalized radius ($\rn$). It is defined by
\begin{equation} \label{NAVdef}
   \delta V^2(\rn) \equiv \frac{\Delta V^2(\rn \, r_\mscript{max})}{\Delta V^2(r_\mscript{max})} \,  , 
\end{equation}
where $r_\mscript{max}$ is the largest radial value with RC data and $\rn \equiv r / r_\mscript{max}$. It is relevant to stress that $r_\mscript{max}$ is not intended to reflect a  physical property of a galaxy, it is simply the radius where the RC observation stops. This end depends on the quality of the observational data. Nonetheless, this parameter is important for studying galaxies of a given sample. In \cite{Rodrigues:2022oyd} we comment on the use of different radial parameters to generate a normalised radius.

From the NAV definition \eqref{NAVdef}, and using the SPARC data \cite{2016AJ....152..157L} for the galaxies without a relevant bulge, we find the NAV observational distribution as given by Fig.~\ref{fig:plotNAVbulgelessObs}.

The essence of the method is to compare the observational NAV distribution $\delta V^2_\mscript{obs}$ with the model-inferred one, which is here due to Palatini gravity and denoted by $\delta V^2_\mscript{P}$. This is significantly easier to do than performing hundreds of fits of galaxies, and moreover, this approach does not require precise knowledge of the baryonic matter distribution: approximations that are not expected to be suitable for all the individual galaxies can be used for the sample, without impact on the $\delta V^2$ sample distribution. This will be done in detail in the next subsection.

\subsection{Baryonic matter approximations} \label{sec:approximations}

For the majority of modified gravity theories, one needs to know the three-dimensional baryonic matter distribution $\rho(r,z) = \rho_*(r,z)+\rho_{\rm gas}(r,z)$. The first difficulty is that these data are not known in detail or provided, and the second is that, for several cases, technically it is significantly harder to derive the rotation curve from modified gravity models than from Newtonian gravity. Hence, it is common that, to test modified gravity, one considers baryonic matter approximations (indeed, \cite{Naik:2018mtx, OBrien:2018znl, Green:2019cqm} are some examples that use SPARC data approximations). We remark that the three-dimensional baryonic matter profile is not needed to test dark matter profiles, the stellar and gaseous contributions to the circular velocity profiles are sufficient for the latter, and these are commonly provided in galaxy rotation curve samples. 

For the stellar component, we use an exponential thin disk as an approximation  \cite{1970ApJ...160..811F, vanderKruit:2011vt}. The surface density is given by
\begin{equation} \label{exponentialDensity}
	\Sigma(r) = \Sigma_0 e^{- r/h} \, ,
\end{equation}
where the constants $h$ and $\Sigma_0$ are found for each one of the 122 galaxies such that their corresponding rotation curve best approximates the stellar rotation curve provided by SPARC (a similar procedure was also done in \cite{Naik:2019moz}, but for a subset of galaxies). The values we derive are different from references that use the exponential profile to extend the observed stellar profile.

The circular velocity due to an axisymmetric exponential profile of negligible thickness has a well-known analytical expression \cite{1970ApJ...160..811F, 0691084459}, which reads
\begin{equation} \label{vExpDisk}
	V^2 = 4 \pi G \Sigma_0  h y^2 [I_0(y) K_0(y) - I_1(y) K_1(y)] \, ,
\end{equation}
where $y \equiv r/(2 h)$, and $I_n$ and $K_n$ are modified Bessel functions of the first and the second kind, respectively. Therefore, we find the stellar exponential component parameters ($h$, $\Sigma_0$) that best approximate the stellar contribution to the SPARC circular velocity by fitting \eqref{vExpDisk} to the latter data.

Although the stellar component has a well-established, simple and good approximation \cite{1970ApJ...160..811F, vanderKruit:2011vt}, the gas component is known to be significantly more diverse.  For instance, \cite{Green:2019cqm} considered elaborated morphology-dependent approximations for the gas; but it also introduced systematical deviations with respect to SPARC data. Reference \cite{Naik:2018mtx} uses an exponential approximation for the gas component, considering a subset of the SPARC data. Contrary to the previously cited approaches, here we do not consider individual galaxy fits, hence the issue of whether such approximations are, or are not, good for specific galaxies is not relevant. This since our focus here is on the sample data distribution, not on the individual particularities. Thus, similarly to \cite{Naik:2018mtx}, we will consider exponential approximations for the gas in the same way we considered the stellar part. Each of the 122 galaxies will be associated with four numbers that describe the stellar and gaseous components (two pairs of $\Sigma_0$ and $h$). We will detail further these approximations in a future work more specific to this subject, but this description is sufficient to reproduce our results.

\subsection{Application}

From eq.~\eqref{V2Palatini} with $\kappa = 8\pi G$, the expressions of $\Delta \Vmod$ and $\delta \Vmod$ for Palatini $f(R)$ read
\begin{equation} \label{DeltaV2Palatini}
	\Delta V^2_\mscript{P}(r) =  8 \pi G \alpha r \rho'(r)  \, ,
\end{equation}
hence
\begin{equation} \label{palatinideltav}
	\delta V^2_\mscript{P}(\rn) =  \rn \frac{ \rho'(\rn)}{\rho'(1)} \, . 
\end{equation}

\begin{figure}
    \centering
	\begin{tikzpicture}
  		\node (img1)  {\includegraphics[width=0.70\textwidth] {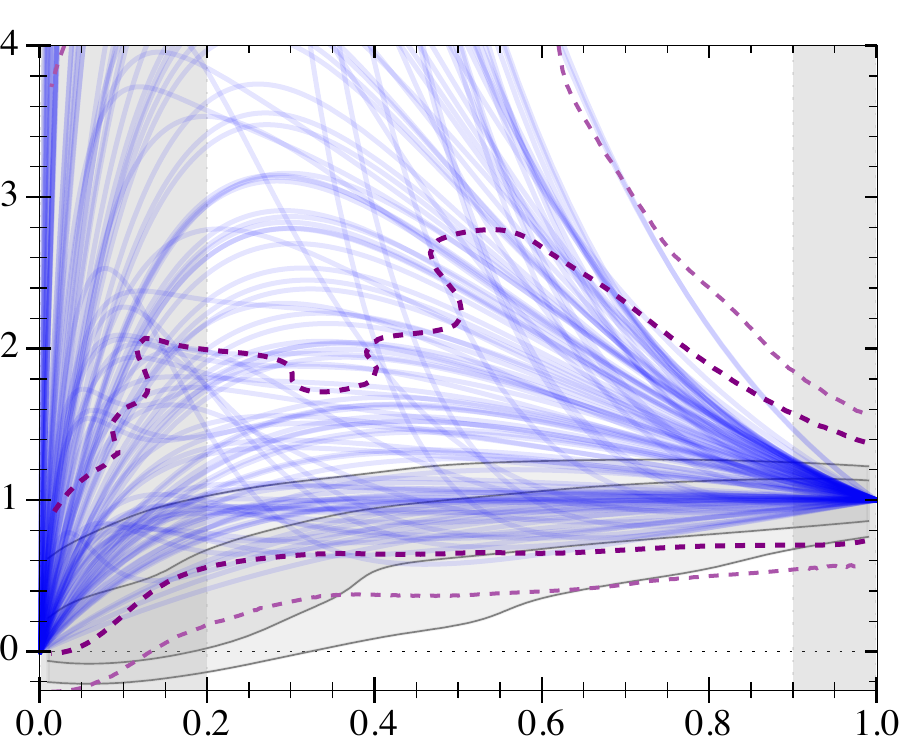}};
		\node[below=of img1, node distance=0cm, yshift=1.1cm, xshift=0.3cm, font=\color{black}] {\large $\rn$};
		\node[left=of img1, node distance=0cm, rotate=90, yshift=-0.9cm, xshift=0.6cm] {\large $\delta V^2$};
	\end{tikzpicture}
	\caption{The NAV plane for $f(R)$ Palatini gravity contrasted with the observational NAV plane. The grey regions are explained in Fig. \ref{fig:plotNAVbulgelessObs}. Each  blue curve corresponds to one of the 122 bulgeless SPARC galaxies. These curves are generated using the approximations of Sec.~\ref{sec:approximations} together with the $\delta V^2_\mscript{P}$ expression \eqref{palatinideltav}. The limits of the 1$\sigma$ and 2$\sigma$ highest density regions for Palatini are shown with the dashed purple contours (dark purple and light purple for the 1$\sigma$ and 2$\sigma$ regions respectively). These model regions are significantly different from the corresponding observational ones. The entire $2\sigma$ Palatini region is not displayed as it extends far beyond the observational region.}
	\label{fig:plotdeltaVPalatini} 
\end{figure}

It is important to stress that $\delta V^2_\mscript{P}$ is independent of $\alpha$. The NAV tests the shape of the additional contribution to the rotation curve, being independent of the amplitude. 

Since $\delta V^2_\mscript{P}$ directly depends on the baryonic matter distribution, we use the approximation detailed in Sec.~\ref{sec:approximations} to generate the Palatini NAV plane and contrast it with the observational NAV, as done in Fig.~\ref{fig:plotdeltaVPalatini}.

From Fig.~\ref{fig:plotdeltaVPalatini}, one sees that Palatini has difficulties on covering the bottom part of the observational NAV region, while there is a large excess of data outside the observational NAV region. It is important to stress that these results are for any $\alpha$ values, which may or may not change from galaxy to galaxy.

In \cite{Rodrigues:2022oyd} we introduce a metric to quantify the concordance between a model and the observational NAV, the mean efficiency $E_\mscript{M}$. It is such that the maximum concordance is $E_\mscript{M} = 1$ and a model whose half of its predicted data is inside the observational region and half of it is outside yields $E_\mscript{M} = 0$. There is no lower limit for $E_\mscript{M}$: the larger the predicted model region outside the observational NAV region, the lower will be $E_\mscript{M}$. Thus good or acceptable models should yield $E_\mscript{M} \gtrsim 0.5$, while the unsuited models would yield $E_\mscript{M} \lesssim 0$. For instance, the well-known dark matter and modified gravity models called Burkert, NFW and MOND yield respectively 0.80, 0.70 and 0.53 for the mean efficiency \cite{Rodrigues:2022oyd}. For $f(R)$ Palatini, we infer that 
\begin{equation}
  E_\mscript{M} < -2.0 \, . \label{meanEff}
\end{equation}

We conclude that Palatini $f(R)$ gravity, or any other theory that satisfies eq.~\eqref{palatinideltav} is \textit{not} compatible with the observational data. That is, these theories cannot replace dark matter in galaxies (even if $\alpha$ changes from galaxy to galaxy).


\section{Conclusions}

Here we considered  Palatini $f(R)$ gravity in different physical scenarios. We show that, although deviations from general relativity (GR) do not appear in the Will Nordtvedt parametrized post-Newtonian (PPN) formalism \cite{Toniato:2019rrd}, it can be constrained within other physical systems. In particular, we revised a recent approach that constrains modified gravity with Earth seismology data \cite{dziewonski1981preliminary}. Our main result refers to testing Palatini gravity as a possible replacement of dark matter in galaxies, using SPARC \cite{2016AJ....152..157L} data for rotationally supported galaxies. 

We have found that, although it is reasonable to use a thin disk approximation for Palatini, replacing dark matter with Palatini gravity leads to a strong tension between the model and the observational data. Indeed, most of its predictions are outside and far from the observational data, as shown in Fig. \ref{fig:plotdeltaVPalatini} and computed in eq.~\eqref{meanEff}. This result is valid for any value of $\alpha$ \eqref{polatiniPotential} (it could be either a common $\alpha$ value for all the galaxies or one that changes from galaxy to galaxy). There are clearly other models that are better for rotation curve data, either within the dark matter approach (e.g., NFW or Burkert halos) or modified gravity (e.g., MOND) \cite{Rodrigues:2022oyd}. 

From eq.~\eqref{polatiniPotential} and simple approximations for the baryonic part contribution, it is not hard to conclude that Palatini $f(R)$ will have difficulties in trying to replace dark matter in galaxies. This is so since dark matter is commonly expected to play a major role in the outskirts of galaxies. In these regions, the baryonic density is small and (apart from individual peculiarities) should become even smaller with increasing radius. On the other hand, the contribution $\nabla^2 \rho$ is expected to be more relevant close to the galaxy centre than far from it. This is in qualitative agreement with our finds here. Nonetheless, this type of qualitative argument, although useful, cannot quantify the discrepancy between the model and data, nor it is useful to compare results from different models.

At last, we stress that the normalised additional velocity (NAV) method \cite{Rodrigues:2022oyd} is a recent one, and we understand that its application to $f(R)$ Palatini is a relevant illustration of how it works. The main code of the method, with several applications, is available at \cite{NAVanalysis}.

\section*{Acknowledgments}

We gratefully acknowledge the organizers of the \textit{Metric-Affine Frameworks for Gravity 2022} workshop for their outstanding work in creating a friendly atmosphere for fruitful discussions and collaborations. AHA thanks CAPES (Brazil) for financial support.  DCR thanks CNPq (Brazil) and FAPES (Brazil) for partial financial support (TO 1020/2022, 976/2022, 1081/2022). AW acknowledges financial support from MICINN (Spain) {\it Ayuda Juan de la Cierva - incorporac\'ion} 2020 No. IJC2020-044751-I.

\appendix

\section{}\label{appA}

Note that \eqref{polatiniPotential} is also a weak field regime for the Eddington-inspired Born-Infeld (EiBI) gravity \cite{Banados:2010ix}
\begin{equation}
    \nabla^2\phi =  \frac{\kappa}{2}\Big(\rho+\alpha\nabla^2\rho\Big) ,
\end{equation}
therefore our results, arriving from the planetary seismology and galaxies, will also refer to this proposal. It is so because the Poisson equation similarity in both theories of gravity is not accidental: let us note that in the 1st order approximation, $ \vert R_{\mu\nu} \vert \ll \alpha^{-1}$, EiBI gravity becomes Palatini gravity with a quadratic term \cite{Pani:2012qd}
\begin{equation} \label{eq:quadgrav}
\mathcal{S}[g,\Gamma,\psi_m] \approx \frac{1}{\kappa}\int d^4x\sqrt{g}\left[R-2\Lambda+\frac{\alpha}{2}(R^2-2R_{\mu\nu}R^{\mu\nu})\right],
\end{equation}
where $\Lambda$ is a cosmological constant. As we already know from the PPN analysis, only the quadratic terms are manifested in the non-relativistic regime in Palatini gravity and hence we are dealing with {\it the same} modified Poisson equation. Therefore, the upper bound resulting from the seismic data for the EiBI proposal is $\alpha\lesssim 2\cdot 10^9 \text{m}^2$ \cite{Kozak:2023axy}.

\definecolor{arsenic}{rgb}{0.23, 0.27, 0.29}
\hypersetup{urlcolor = arsenic}


\begin{thebibliography}{10}

\bibitem{Capozziello:2010zz}
V.~Faraoni and S.~Capozziello, \href{http://dx.doi.org/10.1007/978-94-007-0165-6}{``{Beyond Einstein Gravity}''}, (2011) , vol.~170. 170 of {\em Fundamental Theories of Physics}.
\newblock Springer, Dordrecht.
\newblock
\url{http://www.springerlink.com/content/hl1805/#section=801705&page=1}.
\newblock

\bibitem{Capozziello:2011et}
S.~Capozziello and M.~De~Laurentis, ``{Extended Theories of Gravity},'' \href{http://dx.doi.org/10.1016/j.physrep.2011.09.003}{{\em Phys. Rept.} {\bfseries 509} (2011) 167--321},
\href{http://arxiv.org/abs/1108.6266}{{\ttfamily arXiv:1108.6266 [gr-qc]}}.

\bibitem{Olmo:2011uz}
G.~J. Olmo, ``{Palatini Approach to Modified Gravity: f(R) Theories and Beyond},'' \href{http://dx.doi.org/10.1142/S0218271811018925}{{\em Int. J. Mod. Phys.} {\bfseries D20} (2011) 413--462},
\href{http://arxiv.org/abs/1101.3864}{{\ttfamily arXiv:1101.3864 [gr-qc]}}.

\bibitem{Capozziello:2015lza}
S.~Capozziello, T.~Harko, T.~S. Koivisto, F.~S.~N. Lobo, and G.~J. Olmo, ``{Hybrid metric-Palatini gravity},'' \href{http://dx.doi.org/10.3390/universe1020199}{{\em Universe} {\bfseries 1} no.~2, (2015) 199--238},
\href{http://arxiv.org/abs/1508.04641}{{\ttfamily arXiv:1508.04641 [gr-qc]}}.

\bibitem{tamanini2013generalized}
N.~Tamanini and C.~G. Boehmer, ``Generalized hybrid metric-palatini gravity,'' {\em Physical Review D} {\bfseries 87} no.~8, (2013) 084031.

\bibitem{Ferraris:1992dx}
M.~Ferraris, M.~Francaviglia, and I.~Volovich, ``{The Universality of vacuum Einstein equations with cosmological constant},'' \href{http://dx.doi.org/10.1088/0264-9381/11/6/015}{{\em Class. Quant. Grav.} {\bfseries 11} (1994) 1505--1517},
\href{http://arxiv.org/abs/gr-qc/9303007}{{\ttfamily arXiv:gr-qc/9303007 [gr-qc]}}.

\bibitem{Rodrigues:2018ioe}
D.~C. Rodrigues, M.~Galv{\~a}o, and N.~Pinto-Neto, ``{Hamiltonian analysis of General Relativity and extended gravity from the iterative Faddeev-Jackiw symplectic approach},'' \href{http://dx.doi.org/10.1103/PhysRevD.98.104019}{{\em Phys. Rev.} {\bfseries D98} no.~10, (2018) 104019},
\href{http://arxiv.org/abs/1808.06751}{{\ttfamily arXiv:1808.06751 [gr-qc]}}.

\bibitem{TeppaPannia:2016vsb}
F.~A. Teppa~Pannia, F.~Garc\'\i{}a, S.~E. Perez~Bergliaffa, M.~Orellana, and G.~E. Romero, ``{Structure of Compact Stars in R-squared Palatini Gravity},'' \href{http://dx.doi.org/10.1007/s10714-016-2182-7}{{\em Gen. Rel. Grav.} {\bfseries 49} no.~2, (2017) 25}, \href{http://arxiv.org/abs/1607.03508}{{\ttfamily arXiv:1607.03508 [gr-qc]}}.

\bibitem{Wojnar:2017tmy}
A.~Wojnar, ``{On stability of a neutron star system in Palatini gravity},'' \href{http://dx.doi.org/10.1140/epjc/s10052-018-5900-3}{{\em Eur. Phys. J. C} {\bfseries 78} no.~5, (2018) 421}, \href{http://arxiv.org/abs/1712.01943}{{\ttfamily arXiv:1712.01943 [gr-qc]}}.

\bibitem{Toniato:2019rrd}
J.~D. Toniato, D.~C. Rodrigues, and A.~Wojnar, ``{Palatini $f(R)$ gravity in the solar system: post-Newtonian equations of motion and complete PPN parameters},'' \href{http://dx.doi.org/10.1103/PhysRevD.101.064050}{{\em Phys. Rev.} {\bfseries D101} no.~6, (2020) 064050},
\href{http://arxiv.org/abs/1912.12234}{{\ttfamily arXiv:1912.12234 [gr-qc]}}.

\bibitem{jimenez2018born}
J.~B. Jim{\'e}nez, L.~Heisenberg, G.~J. Olmo, and D.~Rubiera-Garcia, ``Born--infeld inspired modifications of gravity,'' {\em Physics Reports} {\bfseries 727} (2018) 1--129.

\bibitem{Wojnar:2022dvo}
A.~Wojnar, ``{Fermi gas and modified gravity},'' \href{http://dx.doi.org/10.1103/PhysRevD.107.044025}{{\em Phys. Rev. D} {\bfseries 107} no.~4, (2023) 044025}, \href{http://arxiv.org/abs/2208.04023}{{\ttfamily arXiv:2208.04023 [gr-qc]}}.

\bibitem{Benito:2021ywe}
M.~Benito and A.~Wojnar, ``{Cooling process of brown dwarfs in Palatini f(R) gravity},'' \href{http://dx.doi.org/10.1103/PhysRevD.103.064032}{{\em Phys. Rev. D} {\bfseries 103} no.~6, (2021) 064032}, \href{http://arxiv.org/abs/2101.02146}{{\ttfamily arXiv:2101.02146 [gr-qc]}}.

\bibitem{Kalita:2022trq}
S.~Kalita, L.~Sarmah, and A.~Wojnar, ``{Metric-affine effects in crystallization processes of white dwarfs},'' \href{http://dx.doi.org/10.1103/PhysRevD.107.044072}{{\em Phys. Rev. D} {\bfseries 107} no.~4, (2023) 044072}, \href{http://arxiv.org/abs/2212.04918}{{\ttfamily arXiv:2212.04918 [gr-qc]}}.

\bibitem{1972gcpa.book.....W}
S.~{Weinberg}, ``{Gravitation and Cosmology: Principles and Applications of the General Theory of Relativity}'', July, 1972.
\newblock John Wiley \& Sons, Inc.

\bibitem{Will:2014kxa}
C.~M. Will, ``{The Confrontation between General Relativity and Experiment},'' \href{http://dx.doi.org/10.12942/lrr-2014-4}{{\em Living Rev.Rel.} {\bfseries 17} (2014) 4},
\href{http://arxiv.org/abs/1403.7377}{{\ttfamily arXiv:1403.7377 [gr-qc]}}.

\bibitem{Kozak:2023axy}
A.~Kozak and A.~Wojnar, ``{Planetary seismology as a test of modified gravity proposals},'' \href{http://arxiv.org/abs/2303.17213}{{\ttfamily arXiv:2303.17213 [gr-qc]}}.

\bibitem{Capozziello:2006ph}
S.~Capozziello, V.~F. Cardone, and A.~Troisi, ``{Low surface brightness galaxies rotation curves in the low energy limit of r**n gravity: no need for dark matter?},'' \href{http://dx.doi.org/10.1111/j.1365-2966.2007.11401.x}{{\em Mon. Not. Roy. Astron. Soc.} {\bfseries 375} (2007) 1423--1440}, \href{http://arxiv.org/abs/astro-ph/0603522}{{\ttfamily arXiv:astro-ph/0603522}}.

\bibitem{Capozziello:2012ie}
S.~Capozziello and M.~De~Laurentis, ``{The dark matter problem from f(R) gravity viewpoint},''
\href{http://dx.doi.org/10.1002/andp.201200109}{{\em Annalen Phys.} {\bfseries 524} (2012) 545--578}.

\bibitem{Salucci:2014oka}
P.~Salucci, C.~F. Martins, and E.~Karukes, ``{$R^n$ gravity is kicking and alive: the cases of Orion and NGC 3198},'' \href{http://dx.doi.org/10.1142/S021827181442005X}{{\em Int. J. Mod. Phys. D} {\bfseries 23} no.~12, (2014) 1442005}, \href{http://arxiv.org/abs/1405.6314}{{\ttfamily arXiv:1405.6314 [astro-ph.GA]}}.

\bibitem{Naik:2019moz}
A.~P. Naik, E.~Puchwein, A.-C. Davis, D.~Sijacki, and H.~Desmond, ``{Constraints on Chameleon f(R)-Gravity from Galaxy Rotation Curves of the SPARC Sample},'' \href{http://dx.doi.org/10.1093/mnras/stz2131}{{\em \mnras} {\bfseries 489} no.~1, (2019) 771--787}, \href{http://arxiv.org/abs/1905.13330}{{\ttfamily arXiv:1905.13330 [astro-ph.CO]}}.

\bibitem{2016AJ....152..157L}
F.~{Lelli}, S.~S. {McGaugh}, and J.~M. {Schombert}, ``{SPARC: Mass Models for 175 Disk Galaxies with Spitzer Photometry and Accurate Rotation Curves},'' \href{http://dx.doi.org/10.3847/0004-6256/152/6/157}{{\em \aj} {\bfseries 152} (Dec., 2016) 157}, \href{http://arxiv.org/abs/1606.09251}{{\ttfamily arXiv:1606.09251}}.

\bibitem{Rodrigues:2022oyd}
D.~C. Rodrigues, A.~Hernandez-Arboleda, and A.~Wojnar, ``{Normalized additional velocity distribution: Testing the radial profile of dark matter halos and MOND},'' \href{http://dx.doi.org/10.1016/j.dark.2023.101230}{{\em Phys. Dark Univ.} {\bfseries 41} (2023) 101230}, \href{http://arxiv.org/abs/2204.03762}{{\ttfamily arXiv:2204.03762 [astro-ph.GA]}}.

\bibitem{Banados:2010ix}
M.~Ba\~{n}ados and P.~G. Ferreira, ``{Eddington's theory of gravity and its progeny},'' \href{http://dx.doi.org/10.1103/PhysRevLett.105.011101}{{\em Phys.Rev.Lett.} {\bfseries 105} (2010) 011101},
\href{http://arxiv.org/abs/1006.1769}{{\ttfamily arXiv:1006.1769 [astro-ph.CO]}}.

\bibitem{Sotiriou:2008rp}
T.~P. Sotiriou and V.~Faraoni, ``{f(R) Theories Of Gravity},'' \href{http://dx.doi.org/10.1103/RevModPhys.82.451}{{\em Rev.Mod.Phys.} {\bfseries 82} (2010) 451--497},
\href{http://arxiv.org/abs/0805.1726}{{\ttfamily arXiv:0805.1726 [gr-qc]}}.

\bibitem{DeFelice:2010aj}
A.~De~Felice and S.~Tsujikawa, ``{f(R) theories},'' \href{http://dx.doi.org/10.12942/lrr-2010-3}{{\em Living Rev.Rel.} {\bfseries 13} (2010) 3},
\href{http://arxiv.org/abs/1002.4928}{{\ttfamily arXiv:1002.4928 [gr-qc]}}.

\bibitem{Wald:1984rg}
R.~M. Wald, \href{http://dx.doi.org/10.7208/chicago/9780226870373.001.0001}{``{General Relativity}''}, (1984) .
\newblock
Chicago, USA: Univ. Pr.
\newblock

\bibitem{Fabris:2020uey}
J.~C. Fabris, O.~F. Piattella, and D.~C. Rodrigues, ``{On Rastall gravity formulation as a $f(R,\mathcal {L}_m)$ and a f(R,~T) theory},'' \href{http://dx.doi.org/10.1140/epjp/s13360-023-03845-1}{{\em Eur. Phys. J. Plus} {\bfseries 138} no.~3, (2023) 232}, \href{http://arxiv.org/abs/2011.10503}{{\ttfamily arXiv:2011.10503 [gr-qc]}}.

\bibitem{kainulainen2007interior}
K.~Kainulainen, V.~Reijonen, and D.~Sunhede, ``Interior spacetimes of stars in palatini f (r) gravity,'' {\em Physical Review D} {\bfseries 76} no.~4, (2007) 043503.

\bibitem{Will:1993ns}
C.~Will, ``{Theory and experiment in gravitational physics}'', (1993) .
\newblock
Cambridge University Press.
\newblock

\bibitem{Olmo:2005zr}
G.~J. Olmo, ``{The Gravity Lagrangian according to solar system experiments},'' \href{http://dx.doi.org/10.1103/PhysRevLett.95.261102}{{\em Phys. Rev. Lett.} {\bfseries 95} (2005) 261102}, \href{http://arxiv.org/abs/gr-qc/0505101}{{\ttfamily arXiv:gr-qc/0505101}}.

\bibitem{poisson_will_2014}
E.~Poisson and C.~M. Will, \href{http://dx.doi.org/10.1017/CBO9781139507486}{``Gravity: Newtonian, post-newtonian, relativistic''}, (2014) .
\newblock Cambridge University Press, Cambridge.

\bibitem{poirier2000introduction}
J.-P. Poirier, ``Introduction to the physics of the earth's interior'', (2000) .
\newblock Cambridge University Press.

\bibitem{dziewonski1981preliminary}
A.~M. Dziewonski and D.~L. Anderson, ``Preliminary reference earth model,'' {\em Physics of the earth and planetary interiors} {\bfseries 25} no.~4, (1981) 297--356.

\bibitem{kustowski2008anisotropic}
B.~Kustowski, G.~Ekstr{\"o}m, and A.~Dziewo{\'n}ski, ``Anisotropic shear-wave velocity structure of the earth's mantle: A global model,'' {\em Journal of Geophysical Research: Solid Earth} {\bfseries 113} no.~B6, (2008) .

\bibitem{merkel2021femtosecond}
S.~Merkel, S.~Hok, C.~Bolme, D.~Rittman, K.~J. Ramos, B.~Morrow, H.~J. Lee, B.~Nagler, E.~Galtier, E.~Granados, {\em et~al.}, ``Femtosecond visualization of hcp-iron strength and plasticity under shock compression,'' {\em Physical review letters} {\bfseries 127} no.~20, (2021) 205501.

\bibitem{donini2019neutrino}
A.~Donini, S.~Palomares-Ruiz, and J.~Salvado, ``Neutrino tomography of earth,'' {\em Nature Physics} {\bfseries 15} no.~1, (2019) 37--40.

\bibitem{winter2007neutrino}
W.~Winter, ``Neutrino tomography—learning about the earth’s interior using the propagation of neutrinos,'' in {\em Neutrino Geophysics: Proceedings of Neutrino Sciences 2005}, pp.~285--307, Springer.
\newblock (2007) .

\bibitem{van2021probing}
V.~Van~Elewyck, J.~Coelho, E.~Kaminski, and L.~Maderer, ``Probing the earth’s interior with neutrinos,'' {\em Europhysics News} {\bfseries 52} no.~1, (2021) 19--21.

\bibitem{Wojnar:2022ttc}
A.~Wojnar, ``{Giant planet formation in Palatini gravity},'' \href{http://dx.doi.org/10.1103/PhysRevD.105.124053}{{\em Phys. Rev. D} {\bfseries 105} no.~12, (2022) 124053}, \href{http://arxiv.org/abs/2203.16260}{{\ttfamily arXiv:2203.16260 [gr-qc]}}.

\bibitem{Wojnar:2021xbr}
A.~Wojnar, ``{Jupiter and jovian exoplanets in Palatini f(R\textasciimacron{}) gravity},'' \href{http://dx.doi.org/10.1103/PhysRevD.104.104058}{{\em Phys. Rev. D} {\bfseries 104} no.~10, (2021) 104058}, \href{http://arxiv.org/abs/2108.13528}{{\ttfamily arXiv:2108.13528 [gr-qc]}}.

\bibitem{Kozak:2021zva}
A.~Kozak and A.~Wojnar, ``{Non-homogeneous exoplanets in metric-affine gravity},'' \href{http://dx.doi.org/10.1142/S0219887822501572}{{\em Int. J. Geom. Meth. Mod. Phys.} {\bfseries 19} no.~Supp01, (2022) 2250157}, \href{http://arxiv.org/abs/2110.15139}{{\ttfamily arXiv:2110.15139 [gr-qc]}}.

\bibitem{Kozak:2021fjy}
A.~Kozak and A.~Wojnar, ``{Interiors of Terrestrial Planets in Metric-Affine Gravity},'' \href{http://dx.doi.org/10.3390/universe8010003}{{\em Universe} {\bfseries 8} no.~1, (2021) 3}, \href{http://arxiv.org/abs/2111.11199}{{\ttfamily arXiv:2111.11199 [gr-qc]}}.

\bibitem{Kozak:2021ghd}
A.~Kozak and A.~Wojnar, ``{Metric-affine gravity effects on terrestrial exoplanet profiles},'' \href{http://dx.doi.org/10.1103/PhysRevD.104.084097}{{\em Phys. Rev. D} {\bfseries 104} no.~8, (2021) 084097}, \href{http://arxiv.org/abs/2106.14219}{{\ttfamily arXiv:2106.14219 [gr-qc]}}.

\bibitem{luzum2011iau}
B.~Luzum, N.~Capitaine, A.~Fienga, W.~Folkner, T.~Fukushima, J.~Hilton, C.~Hohenkerk, G.~Krasinsky, G.~Petit, E.~Pitjeva, {\em et~al.}, ``The iau 2009 system of astronomical constants: the report of the iau working group on numerical standards for fundamental astronomy,'' {\em Celestial Mechanics and Dynamical Astronomy} {\bfseries 110} (2011) 293--304.

\bibitem{chen2015consistent}
W.~Chen, J.~C. Li, J.~Ray, W.~B. Shen, and C.~L. Huang, ``Consistent estimates of the dynamic figure parameters of the earth,'' {\em Journal of Geodesy} {\bfseries 89} (2015) 179--188.

\bibitem{olmo2005gravity}
G.~J. Olmo, ``The gravity lagrangian according to solar system experiments,'' {\em Physical review letters} {\bfseries 95} no.~26, (2005) 261102.

\bibitem{banerjee2017constraints}
S.~Banerjee, S.~Shankar, and T.~P. Singh, ``Constraints on modified gravity models from white dwarfs,'' {\em Journal of Cosmology and Astroparticle Physics} {\bfseries 2017} no.~10, (2017) 004.

\bibitem{0691084459}
J.~Binney and S.~Tremaine, ``{Galactic Dynamics (Princeton Series in Astrophysics)}'', (1988) .
\newblock Princeton University Press.

\bibitem{McGaugh:2006vv}
S.~S. McGaugh, W.~J.~G. de~Blok, J.~M. Schombert, R.~K. de~Naray, and J.~H. Kim, ``{The Rotation Velocity Attributable to Dark Matter at Intermediate Radii in Disk Galaxies},'' \href{http://dx.doi.org/10.1086/511807}{{\em Astrophys. J.} {\bfseries 659} (2007) 149--161}, \href{http://arxiv.org/abs/astro-ph/0612410}{{\ttfamily arXiv:astro-ph/0612410}}.

\bibitem{McGaugh:2016leg}
S.~McGaugh, F.~Lelli, and J.~Schombert, ``{Radial Acceleration Relation in Rotationally Supported Galaxies},'' \href{http://dx.doi.org/10.1103/PhysRevLett.117.201101}{{\em Phys. Rev. Lett.} {\bfseries 117} no.~20, (2016) 201101},
\href{http://arxiv.org/abs/1609.05917}{{\ttfamily arXiv:1609.05917 [astro-ph.GA]}}.

\bibitem{Naik:2018mtx}
A.~P. Naik, E.~Puchwein, A.-C. Davis, and C.~Arnold, ``{Imprints of Chameleon f(R) Gravity on Galaxy Rotation Curves},'' \href{http://dx.doi.org/10.1093/mnras/sty2199}{{\em Mon. Not. Roy. Astron. Soc.} {\bfseries 480} no.~4, (2018) 5211--5225}, \href{http://arxiv.org/abs/1805.12221}{{\ttfamily arXiv:1805.12221 [astro-ph.CO]}}.

\bibitem{OBrien:2018znl}
J.~G. O'Brien, T.~L. Chiarelli, P.~D. Mannheim, M.~A. Falcone, M.~H. AlQurashi, and J.~Carter, ``{Radial Acceleration and Tully-Fisher Relations in Conformal Gravity},'' \href{http://dx.doi.org/10.1088/1742-6596/1239/1/012009}{{\em J. Phys. Conf. Ser.} {\bfseries 1239} no.~1, (2019) 012009}, \href{http://arxiv.org/abs/1812.03152}{{\ttfamily arXiv:1812.03152 [astro-ph.GA]}}.

\bibitem{Green:2019cqm}
M.~A. Green and J.~W. Moffat, ``{Modified Gravity (MOG) fits to observed radial acceleration of SPARC galaxies},'' \href{http://dx.doi.org/10.1016/j.dark.2019.100323}{{\em Phys. Dark Univ.} {\bfseries 25} (2019) 100323},
\href{http://arxiv.org/abs/1905.09476}{{\ttfamily arXiv:1905.09476 [gr-qc]}}.

\bibitem{1970ApJ...160..811F}
K.~C. {Freeman}, ``{On the Disks of Spiral and so Galaxies},'' \href{http://dx.doi.org/10.1086/150474}{{\em \apj} {\bfseries 160} (June, 1970) 811}.

\bibitem{vanderKruit:2011vt}
P.~C. van~der Kruit and K.~C. Freeman, ``{Galaxy Disks},'' \href{http://dx.doi.org/10.1146/annurev-astro-083109-153241}{{\em Ann. Rev. Astron. Astrophys.} {\bfseries 49} (2011) 301--371},
\href{http://arxiv.org/abs/1101.1771}{{\ttfamily arXiv:1101.1771 [astro-ph.GA]}}.

\bibitem{NAVanalysis}
D.~C. Rodrigues and A.~Hernandez-Arboleda, ``{NAV}analysis: {N}ormalized {A}dditional {V}elocity distribution analysis,'' \href{http://dx.doi.org/10.5281/zenodo.7373468}{{\em Zenodo} (Nov., 2022) }. \url{https://github.com/cosmo-ufes/NAVanalysis}.

\bibitem{Pani:2012qd}
P.~Pani and T.~P. Sotiriou, ``{Surface singularities in Eddington-inspired Born-Infeld gravity},'' \href{http://dx.doi.org/10.1103/PhysRevLett.109.251102}{{\em Phys. Rev. Lett.} {\bfseries 109} (2012) 251102}, \href{http://arxiv.org/abs/1209.2972}{{\ttfamily arXiv:1209.2972 [gr-qc]}}.

\end{thebibliography}

\providecommand{\href}[2]{#2}\begingroup\raggedright\endgroup

\end{document}